\documentclass[10pt,twocolumn]{article}

\usepackage[a4paper,top=1.55cm,bottom=1.55cm,left=1.55cm,right=1.55cm,columnsep=0.55cm]{geometry}
\usepackage[T1]{fontenc}
\usepackage[utf8]{inputenc}
\usepackage{newtxtext,newtxmath}
\usepackage{microtype}
\usepackage{graphicx}
\usepackage{amsmath,bm}
\usepackage{booktabs}
\usepackage{multirow}
\usepackage{array}
\usepackage{siunitx}
\usepackage{enumitem}
\usepackage{xcolor}
\usepackage{cite}
\usepackage[hidelinks]{hyperref}
\usepackage{caption}
\usepackage{subcaption}
\usepackage{placeins}
\usepackage{balance}

\setlist{nosep,leftmargin=*}
\title{\bfseries Time-Integrated Leakage as a Dynamical Benchmark for Multi-Mode Superconducting-Qubit Reset}

\author{
Sushant Sharma\\
\small Department of Electronics and Communication\\
\small Indian Institute of Information Technology, Dharwad\\
\small \texttt{24bec062@iiitdwd.ac.in}
\and
Aswath Babu H\\
\small Department of Arts, Science and Design\\
\small Indian Institute of Information Technology, Dharwad\\
\small \texttt{aswath@iiitdwd.ac.in}
}
\date{}

\begin{document}
\maketitle
\vspace{-1.2em}

\begin{abstract}
Reset performance in superconducting qubits is often summarized by the population that remains at the end of the pulse. This is an important number, but it does not describe what happens during the reset itself. A protocol may reach a small final residual while keeping the qubit in a leakage state for a comparatively long part of the trajectory. In this work we therefore use time-integrated leakage, $\mathcal{L}_f=\int P_f(t)\,dt$, together with the usual endpoint population to study a multi-mode dissipative reset model under flux-frequency control. We examine finite thermal occupation, parameter sweeps, frequency disorder, an auxiliary-chain ablation, and a literature-based rate benchmark. At the representative ablation point, adding the auxiliary chain reduces $\mathcal{L}_f$ from $8.56$ to $0.78\,$ns and shifts the sustained $P_f<10^{-2}$ crossing from $39.5$ to $9.8\,$ns. Interestingly, the configuration without the chain still reaches the smaller long-time residual. The two metrics can therefore favor different reset configurations. For the same $300\,$ns analysis window, an effective rate model built from the $\lvert f\rangle$ and $\lvert e\rangle$ decay rates reported by Zhou \emph{et al.} gives $\mathcal{L}_f=108.0\,$ns, while the complete simulated configuration gives $0.783\,$ns. We use this only as a kinetic reference and not as a reproduction of the experiment. In the wider set of simulations, thermal occupation is the main source of degradation, while the frequency disorder considered here has a much weaker effect. The results suggest that time-integrated leakage is a useful quantity to report alongside endpoint residuals, especially when reset is part of repeated quantum-error-correction cycles.
\end{abstract}


\section{Introduction}

As superconducting processors are pushed toward repeated error-correction cycles, reset becomes more than a simple preparation step between experiments. Population that leaks into higher transmon levels can survive into later operations and disturb subsequent syndrome measurements~\cite{Acharya2024,McEwen2021,Miao2023}. For a three-level transmon, this means that the $\lvert f\rangle$ population matters not only after reset has finished, but also while the reset is taking place.

A number of reset approaches have already been demonstrated. Measurement-based active reset can produce low residual excitation, although it comes with the additional cost of measurement and feedback~\cite{Han2023,Magnard2018}. Parametric and dissipative schemes avoid that feedback loop and may operate faster, but their performance can be sensitive to detuning, linewidth, thermal occupation, and calibration~\cite{Zhou2021,Sunada2022,Yuan2023}. Multi-mode Purcell structures offer a related idea: instead of depending on one narrow resonance, the dissipative response is distributed over several modes~\cite{Gu2025}.

Reset performance is commonly reported through the population remaining after the pulse. We retain that quantity, but also look at the area under the leakage-population curve. Two protocols may finish with similar residual populations and still spend very different amounts of time in the leakage manifold. We therefore define the \emph{time-integrated leakage}
\begin{equation}
\mathcal{L}_{f}=\int_{0}^{t_f} P_f(t)\,dt,
\label{eq:intleak}
\end{equation}
where $P_f(t)$ is the instantaneous population of $\lvert f\rangle$. The quantity gives a measure of the cumulative leakage exposure during reset, which can be relevant when the same reset operation is repeated many times inside a QEC cycle~\cite{Lacroix2023,Gao2025}. To place the metric on an external timescale, we also construct a simple effective-rate model from the two-tone $\lvert f\rangle$ reset kinetics reported by Zhou \emph{et al.}~\cite{Zhou2021}. This comparison is deliberately limited: it is not meant to reproduce their full device or pulse sequence.

\subsection{Multi-mode dissipation with flux control}

The numerical model combines time-dependent qubit-frequency modulation with a chain of coupled auxiliary modes and an additional lossy mode. The aim is to provide more than one route for excitation to leave the transmon instead of relying on a single sharply tuned channel.

\begin{figure}[t]
    \centering
    \includegraphics[width=\columnwidth]{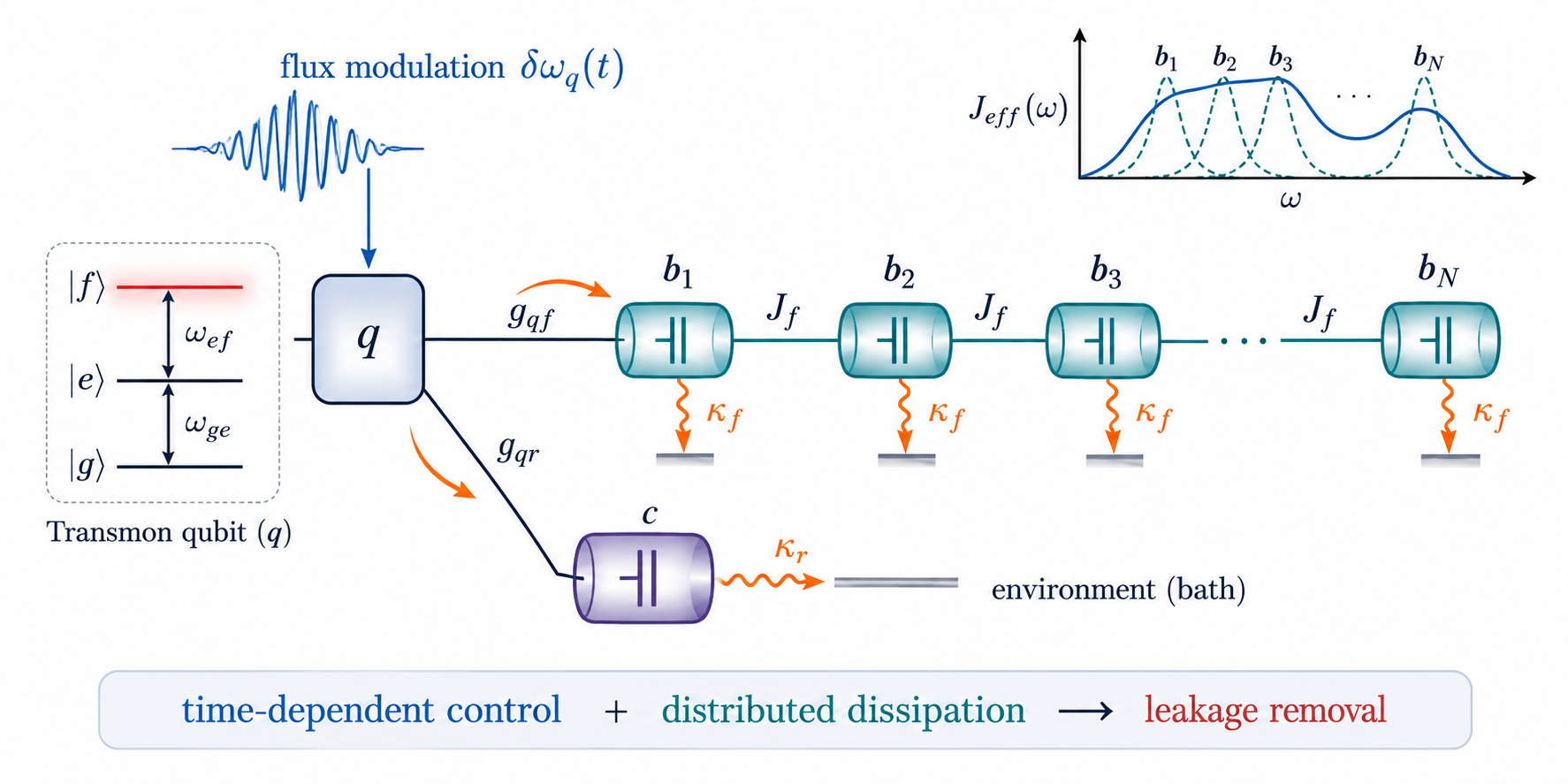}
    \caption{Architecture studied in this work. Time-dependent flux modulation shifts the transmon frequency, the coupled auxiliary-mode chain provides distributed dissipative pathways, and the lossy resonator supplies an additional excitation-removal channel.}
    \label{fig:overview}
\end{figure}

\paragraph{Temporal control through flux modulation.}
In the implementation used for the parameter sweeps, the modulation enters as a time-dependent shift of the transmon frequency. This gives the reset an explicitly time-dependent control parameter without introducing a separate measurement-and-feedback loop.

\paragraph{Coupled auxiliary-mode chain.}
The transmon couples directly to the first auxiliary mode, while neighboring auxiliary modes are coupled to one another. The chain therefore provides a distributed set of frequencies and couplings rather than a single isolated resonance. In the sweeps we vary the number of auxiliary modes between three and seven.

\paragraph{Dissipative channels.}
Loss channels are included phenomenologically through Lindblad collapse operators. The numerical study is therefore intended as a comparative open-system model: the conclusions below refer to trends within this implementation rather than to a device-calibrated experimental model.

\subsection{Simulation program and main contributions}

The numerical study is built around four practical questions. First, how much does finite thermal occupation limit the reset? Second, how do resonator loss and auxiliary-chain size change the endpoint residual and the integrated leakage? Third, how sensitive are the results to random shifts of the auxiliary-mode frequencies? Finally, at one representative operating point, which part of the architecture is mainly responsible for the short-time leakage dynamics? The main observations from the model are:
\begin{enumerate}[label=(\arabic*)]
    \item best-case vacuum simulations reach residual $\lvert f\rangle$ populations of order $10^{-5}$;
    \item finite-temperature sweeps identify the filter thermal occupation as the dominant predictor of residual leakage;
    \item the tested high-loss range and a five-mode auxiliary chain provide a favorable operating region within the parameter sweep;
    \item an ablation study shows an approximately eleven-fold reduction in integrated leakage when the auxiliary chain is enabled, even though the chain-free configuration reaches a lower asymptotic $P_f$;
    \item sweeping the flux-modulation amplitude from $0$ to $1\,\mathrm{ns}^{-1}$ changes integrated leakage by less than $0.6\%$ at the representative point, so the present data do not support a separate modulation-induced enhancement there;
    \item a literature-derived effective-rate benchmark gives $\mathcal{L}_f=107.99\,$ns over the same $300\,$ns window, compared with $0.783\,$ns for the complete configuration; and
    \item the disorder study shows weak sensitivity to random filter detuning up to $\pm60\,$MHz in the simulated device model.
\end{enumerate}

\section{System Model and Methods}

\subsection{Hamiltonian}

We model the transmon with three levels and write its static Hamiltonian as a truncated Duffing oscillator. The auxiliary elements are represented by two-level truncated modes in the numerical implementation. The ordering used in the code is the transmon, followed by the auxiliary chain, followed by the resonator. The static Hamiltonian corresponding to the parameter-sweep implementation is
\begin{align}
\frac{H_0}{\hbar}={}&\omega_{ge}a^{\dagger}a
+\frac{\alpha}{2}a^{\dagger}a^{\dagger}aa
+\sum_{j=1}^{N_f}\omega_{f,j}b_j^{\dagger}b_j
+\omega_r c^{\dagger}c \nonumber\\
&+g_{qf}\left(a^{\dagger}b_1+a b_1^{\dagger}\right)
+J_f\sum_{j=1}^{N_f-1}
\left(b_j^{\dagger}b_{j+1}+b_j b_{j+1}^{\dagger}\right) \nonumber\\
&+g_{qr}\left(a^{\dagger}c+a c^{\dagger}\right).
\label{eq:H0}
\end{align}
Here $a$, $b_j$, and $c$ denote the transmon, the $j$th auxiliary mode, and the resonator operators. For the parameter-sweep code provided with this study,
\begin{equation}
\frac{\omega_{ge}}{2\pi}=4.88~\mathrm{GHz},\qquad
\frac{\alpha}{2\pi}=-0.26~\mathrm{GHz},\qquad
\frac{\omega_r}{2\pi}=6.4~\mathrm{GHz},
\end{equation}
with $g_{qf}/2\pi=0.25~\mathrm{GHz}$, $J_f/2\pi=0.5~\mathrm{GHz}$, and $g_{qr}/2\pi=0.15~\mathrm{GHz}$. The $e\!\leftrightarrow\! f$ transition is therefore centered at
\begin{equation}
\frac{\omega_{ef}}{2\pi}=\frac{\omega_{ge}+\alpha}{2\pi}=4.62~\mathrm{GHz}.
\end{equation}
The auxiliary frequencies are generated around this transition using a nominal $260~\mathrm{MHz}$ spacing parameter according to the numerical rule
\begin{equation}
\omega_{f,j}=\omega_{ef}-2\pi(0.13~\mathrm{GHz})
+2\pi(0.26~\mathrm{GHz})\frac{j-1}{N_f},
\end{equation}
for $j=1,\ldots,N_f$. We state this formula explicitly because it is the one used in the sweep implementation.

The flux modulation is implemented as a time-dependent shift of the transmon number operator,
\begin{equation}
H_{\mathrm{mod}}(t)=\hbar\,\delta\omega_q(t)\,a^{\dagger}a,
\label{eq:drive}
\end{equation}
with
\begin{equation}
\delta\omega_q(t)=
-A_m\exp\!\left[-\frac{(t-t_c)^2}{2\sigma^2}\right]
\cos(\omega_d t),
\end{equation}
for $t\le t_{\mathrm{cut}}$, and zero afterward. In the supplied parameter-sweep implementation, $A_m=0.5~\mathrm{ns}^{-1}$, $\sigma=18~\mathrm{ns}$, $t_c=0.6\sigma$, $t_{\mathrm{cut}}=240~\mathrm{ns}$, and
\begin{equation}
\omega_d=\left|\omega_{ge}-\omega_r+\alpha\right|
=2\pi\times1.78~\mathrm{GHz}.
\end{equation}
This form matches the numerical model directly: the drive is a frequency modulation of the transmon, rather than a transverse $(a+a^{\dagger})$ drive.

\subsection{Open-system dynamics}

The density operator evolves according to the Lindblad master equation
\begin{equation}
\dot{\rho}=-\frac{i}{\hbar}[H_0+H_{\mathrm{mod}}(t),\rho]
+\sum_k\mathcal{D}[L_k]\rho,
\label{eq:lindblad}
\end{equation}
with
\begin{equation}
\mathcal{D}[L]\rho=L\rho L^{\dagger}-\frac{1}{2}\left(L^{\dagger}L\rho+\rho L^{\dagger}L\right).
\end{equation}
The collapse operators in the numerical model describe qubit relaxation and dephasing together with effective loss and finite-temperature excitation channels for the auxiliary degrees of freedom. The parameter-sweep implementation uses $\gamma_1=2\pi\times10^{-5}~\mathrm{ns}^{-1}$ and $\gamma_{\phi}=2\pi\times2.5\times10^{-5}~\mathrm{ns}^{-1}$. The auxiliary and resonator loss parameters are treated as effective phenomenological rates and are varied in the numerical studies below. Simulations are performed using QuTiP's master-equation solver~\cite{Johansson2013} with 1200 time samples over $300\,$ns and initial transmon state $\lvert f\rangle$. Because the dissipative model is effective rather than device-calibrated, the results are interpreted primarily as comparative trends within a common simulation framework.

\subsection{Performance metrics}

We evaluate the final residual population $P_f(t_f)$, the final ground-state population $P_g(t_f)$, and the integrated leakage $\mathcal{L}_f$ in Eq.~\eqref{eq:intleak}. We also record threshold times. For oscillatory trajectories, a first crossing can be misleading because $P_f$ may rise above the same threshold again. In the ablation and external benchmark we therefore use the sustained threshold time
\begin{equation}
t_{\mathrm{s}}(\epsilon)=\min\left\{t:\;P_f(t')<\epsilon\;\text{for all}\;t'\ge t\right\}.
\label{eq:sustained}
\end{equation}
We use ``integrated leakage'' to mean the ensemble-averaged quantity in Eq.~\eqref{eq:intleak}. It has units of time, but it is not a deterministic dwell time for a single realization.

\subsection{Simulation campaigns}

\paragraph{Thermal-noise study.}
We sample $\bar n_{\mathrm{filter}}\in[0.01,0.1]$ and $\bar n_{\mathrm{res}}\in[0.01,0.05]$, with $N_{\mathrm{filter}}=5$ and a fixed baseline resonator-loss setting. These ranges are used to represent imperfect thermalization of the auxiliary microwave degrees of freedom within the effective model.

\paragraph{Parametric optimization.}
We sweep the effective high-loss parameter over the numerical values $\{3,4,5,6,8,10\}$ used in the simulation code, together with $N_{\mathrm{filter}}\in\{3,5,7\}$ and $\bar n_{\mathrm{filter}}\in\{0.015,0.030,0.045,0.060,0.075,0.090\}$. The thermal occupations of the remaining degrees of freedom are held fixed in each sweep. We retain the numerical sweep labels in the Results section so that they can be traced directly to the recorded data.

\paragraph{Fabrication-disorder study.}
For disorder strength $\Delta\in\{10,20,30,40,50,60\}\,$MHz, each filter frequency is shifted as
\begin{equation}
\omega_{f,j}\rightarrow \omega_{f,j}+\delta_{f,j},\qquad
\delta_{f,j}\sim\mathcal{U}(-\Delta,+\Delta),
\end{equation}
with multiple independent realizations per disorder level.

\paragraph{Ablation and flux-amplitude sensitivity.}
To separate endpoint behavior from transient leakage exposure, we perform a controlled ablation at a representative operating point with $N_{\mathrm{filter}}=5$, high-loss setting $8$, $\bar n_{\mathrm{filter}}=0.03$, and $\bar n_{\mathrm{res}}=0.02$. Four configurations are compared while all common parameters are held fixed: a control case with flux modulation and qubit--auxiliary coupling disabled, a flux-only case, a chain-only case, and the complete configuration. The direct qubit--resonator pathway and the common dissipative channels are retained in every case. We then keep the auxiliary chain enabled and sweep the modulation amplitude over $A_m\in\{0,0.25,0.50,0.75,1.00\}\,\mathrm{ns}^{-1}$.

\paragraph{Literature-derived effective-rate benchmark.}
For a focused external comparison, we use the two-tone $\lvert f\rangle$-reset kinetics reported by Zhou \emph{et al.}~\cite{Zhou2021}. Their multi-level decay analysis gives effective rates of approximately $\Gamma_f=(117\,\mathrm{ns})^{-1}$ and $\Gamma_e=(100\,\mathrm{ns})^{-1}$. We represent those fitted rates by
\begin{align}
\dot P_f&=-\Gamma_f P_f,\nonumber\\
\dot P_e&=\Gamma_f P_f-\Gamma_e P_e,\nonumber\\
P_g&=1-P_f-P_e,
\label{eq:ratebenchmark}
\end{align}
with $P_f(0)=1$. The same $300\,$ns analysis window used for the complete configuration is applied to this rate model. This is an effective kinetic benchmark, not a reconstruction of the experimental Hamiltonian, pulse waveform, resonator depletion, thermal floor, or readout model. Since $P_f(t)=e^{-\Gamma_f t}$ in this model, threshold times follow from $t(\epsilon)=\Gamma_f^{-1}\ln(1/\epsilon)$.

\section{Results}

\subsection{Thermal-noise performance}

\begin{figure}[t]
    \centering
    \includegraphics[width=\columnwidth]{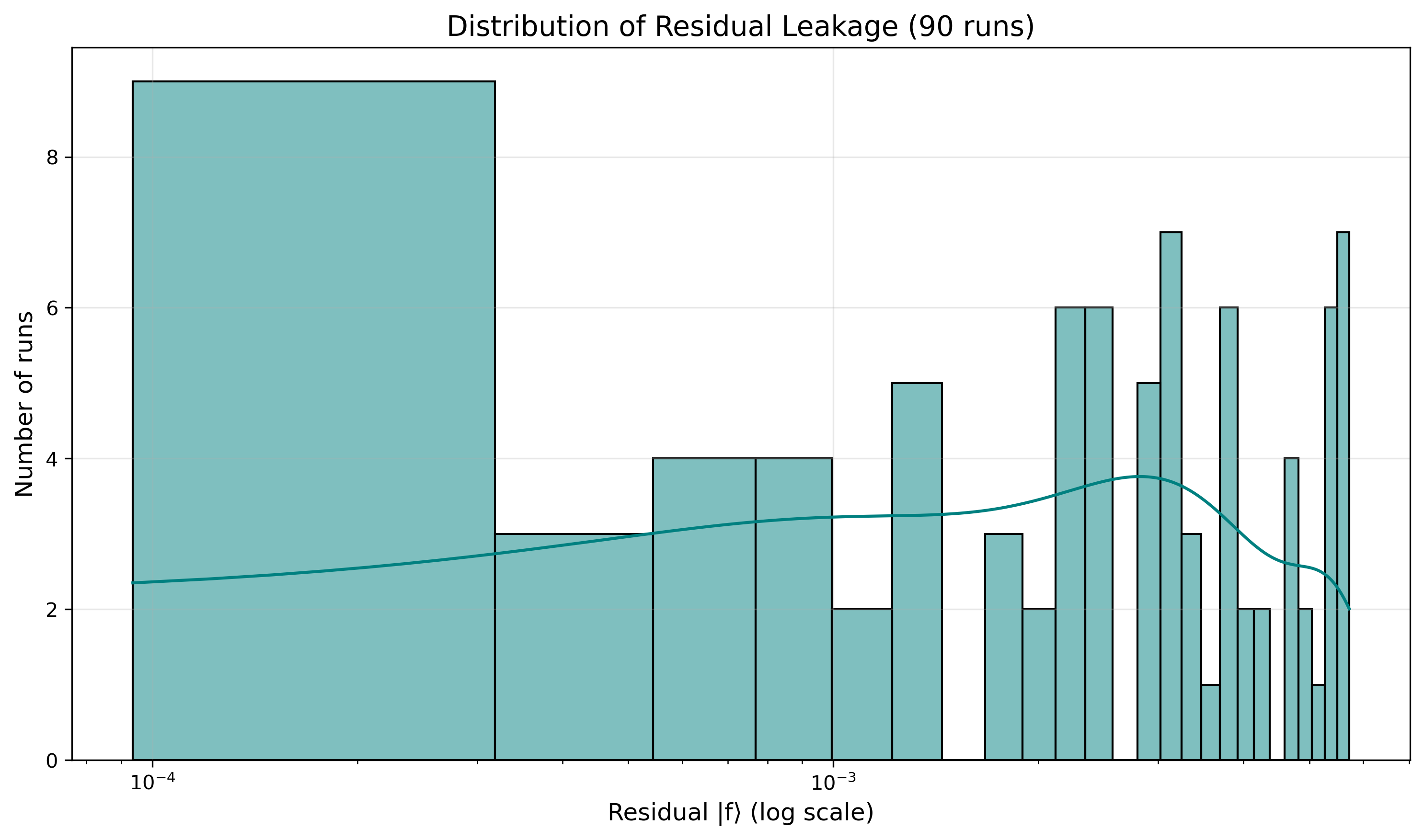}
    \caption{Distribution of residual leakage across the thermal-noise simulations. The broad upper tail reflects thermal re-excitation away from the lowest-residual operating regime.}
    \label{fig:residual_distribution}
\end{figure}

Table~\ref{tab:thermal} summarizes the 90-run thermal-noise data set. The mean residual is $0.236\%$ with a standard deviation of $0.121\%$, while the mean integrated leakage is approximately $1.15\,$ns. Because the recorded summary used for this manuscript does not provide a directly verified standard deviation for the integrated-leakage column, that entry is left unreported rather than inferred. The distribution in Fig.~\ref{fig:residual_distribution} shows that thermal re-excitation produces a broad tail away from the lowest-leakage regime.

\begin{table}[t]
\centering
\caption{Summary statistics from 90 thermal-noise simulations.}
\label{tab:thermal}
\small
\begin{tabular}{lccc}
\toprule
Metric & Mean & Std. dev. & Max. \\
\midrule
Residual $\lvert f\rangle$ & $0.236\%$ & $0.121\%$ & $0.513\%$ \\
Final $\lvert g\rangle$ & $94.8\%$ & $1.9\%$ & $97.4\%$ \\
Final $\lvert e\rangle$ & $4.9\%$ & $1.9\%$ & --- \\
Integrated leakage (ns) & $1.15$ & --- & $1.32$ \\
$t$ to $P_f<10^{-2}$ & $63\%$ reached & --- & --- \\
$t$ to $P_f<10^{-3}$ & $32\%$ reached & --- & --- \\
$t$ to $P_f<10^{-4}$ & $4.6\%$ reached & --- & --- \\
\bottomrule
\end{tabular}
\end{table}

\begin{figure}[t]
    \centering
    \includegraphics[width=\columnwidth]{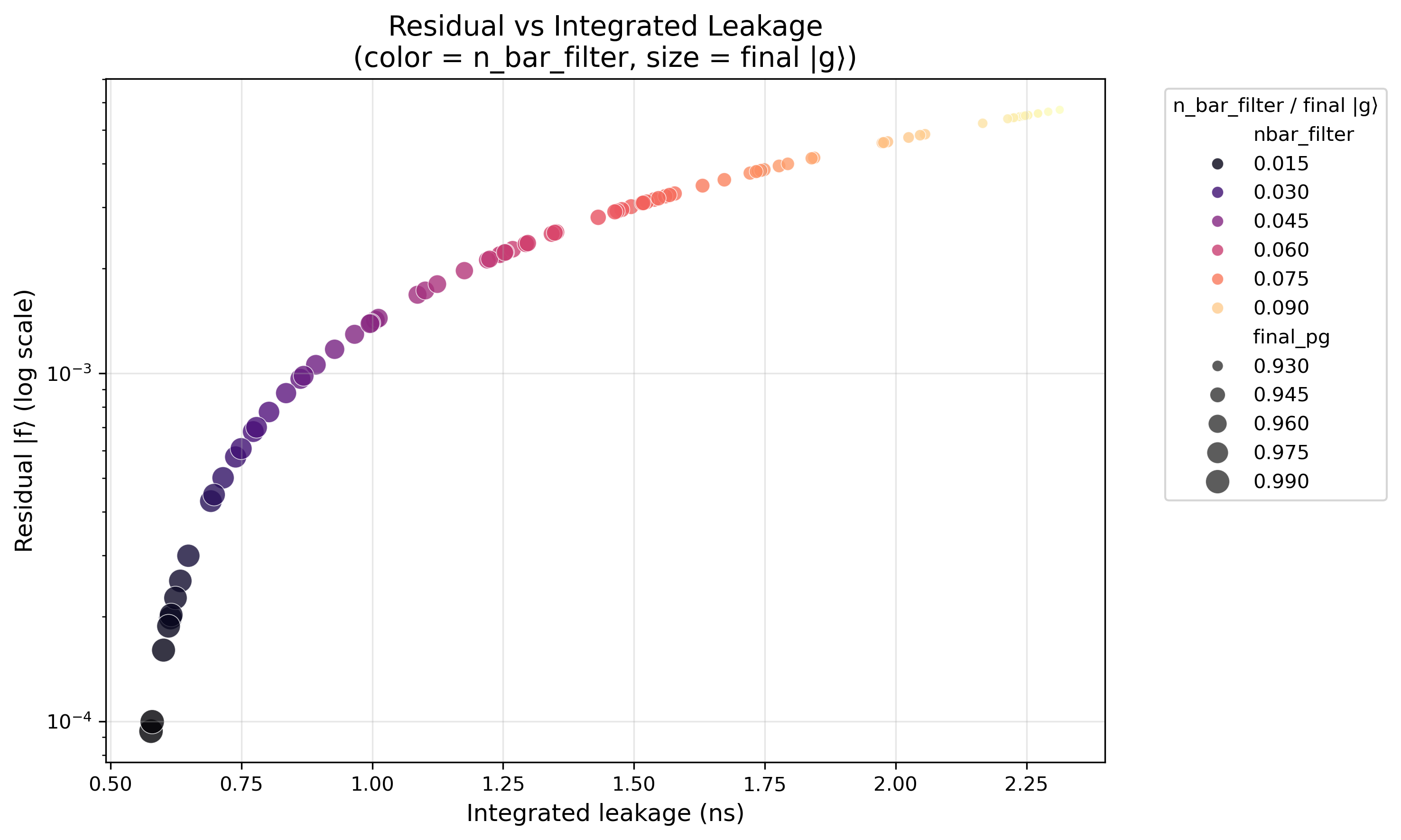}
    \caption{Residual population versus integrated leakage. Point color denotes filter thermal occupation and point size denotes final ground-state population.}
    \label{fig:resid_int}
\end{figure}

Figure~\ref{fig:resid_int} compares residual population against integrated leakage. Both quantities increase together as thermal occupation rises, showing that the thermal environment affects not only the endpoint of the reset but also the full transient trajectory.

\begin{figure}[t]
    \centering
    \includegraphics[width=\columnwidth]{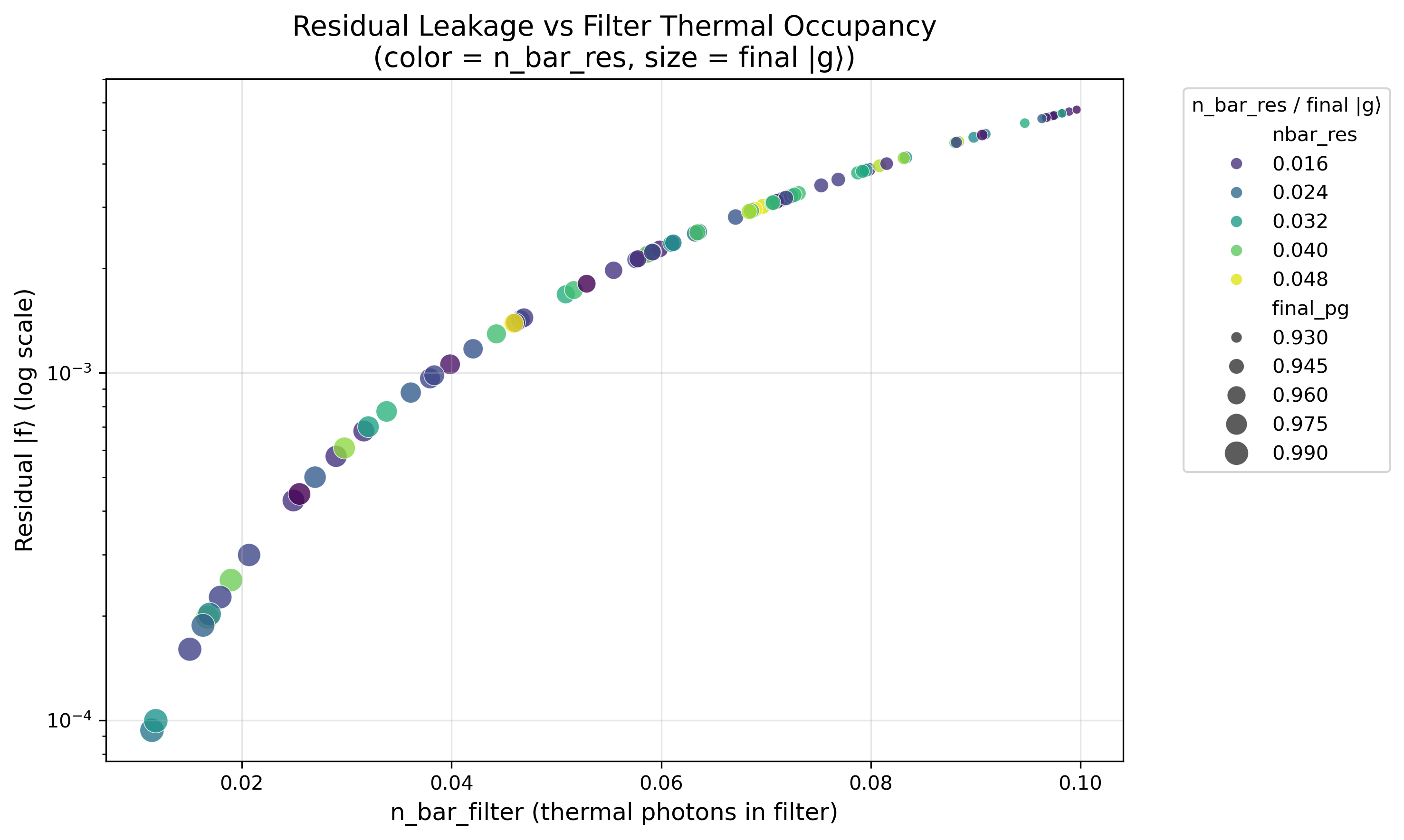}
    \caption{Residual leakage versus filter thermal occupation. The monotonic trend indicates that filter thermal photons dominate the variation across the sampled thermal-noise data.}
    \label{fig:thermalcorr}
\end{figure}

The filter thermal occupation exhibits the strongest correlation with residual leakage in the sampled data, with $R^2>0.95$ in the regression analysis of the thermal-noise data. Figure~\ref{fig:thermalcorr} therefore identifies filter thermalization as the most important practical limitation within the present model.

\subsection{Parametric optimization}

The parameter sweep shows a clear dependence on resonator loss and a weaker but non-monotonic dependence on filter-chain size. The main summary is given in Table~\ref{tab:parametric}.

\begin{table}[t]
\centering
\caption{Parametric optimization summary.}
\label{tab:parametric}
\small
\begin{tabular}{lcc}
\toprule
Configuration & Mean residual & Int. leakage (ns) \\
\midrule
All 69 runs & $0.163\%$ & $1.14$ \\
loss parameter $=3$--$4$ & $0.274\%$ & $1.32$ \\
loss parameter $=8$--$10$ & $0.164\%$ & $1.12$ \\
$N_{\mathrm{filter}}=3$ & $0.189\%$ & $1.21$ \\
$N_{\mathrm{filter}}=5$ & $0.162\%$ & $1.14$ \\
$\bar n_{\mathrm{filter}}=0.015$ & $0.089\%$ & $0.92$ \\
$\bar n_{\mathrm{filter}}=0.090$ & $0.312\%$ & $1.54$ \\
\bottomrule
\end{tabular}
\end{table}

\begin{figure}[t]
    \centering
    \includegraphics[width=\columnwidth]{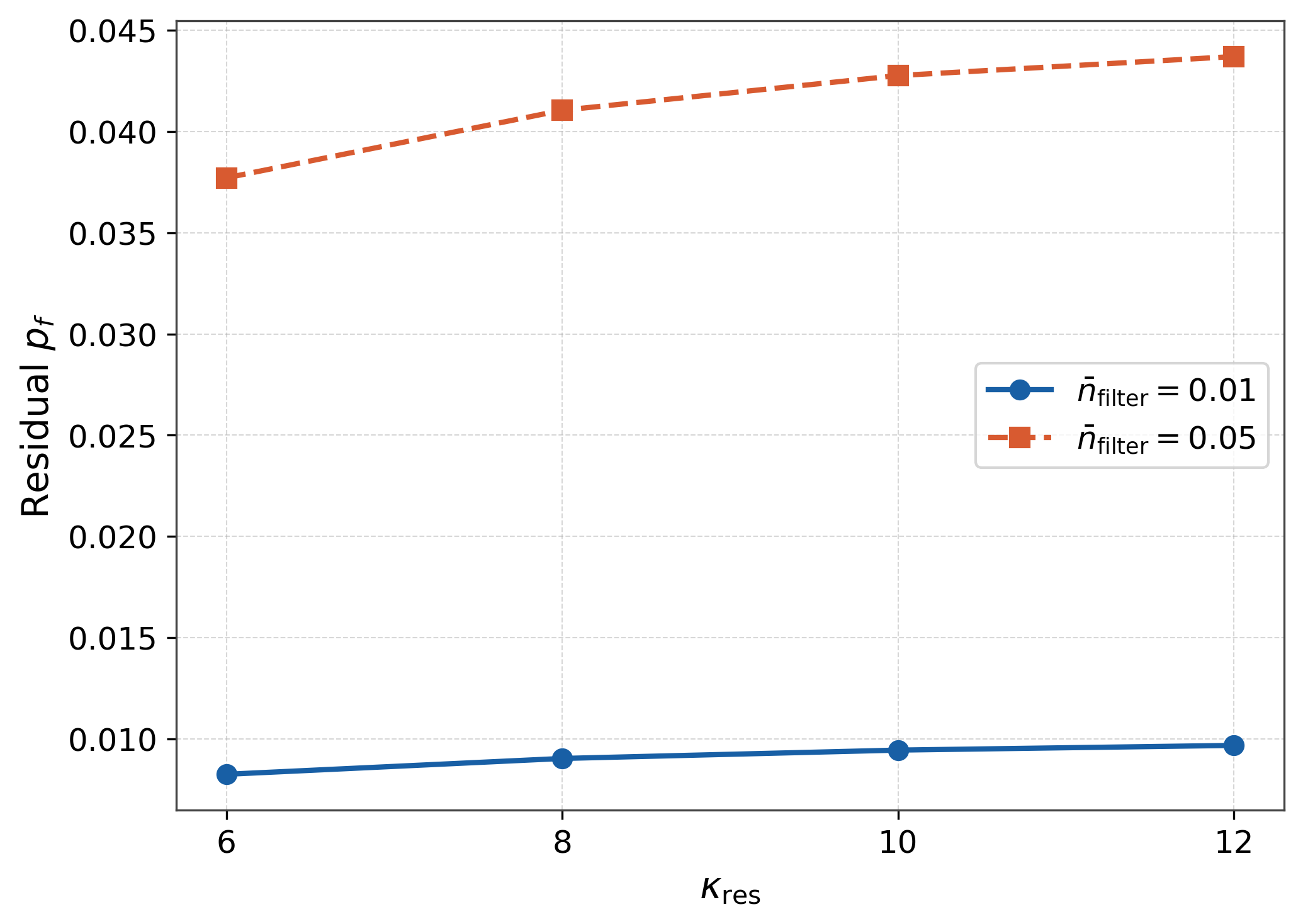}
    \caption{Residual population as a function of resonator loss for two representative filter thermal occupations.}
    \label{fig:reskappa}
\end{figure}

\begin{figure}[t]
    \centering
    \includegraphics[width=\columnwidth]{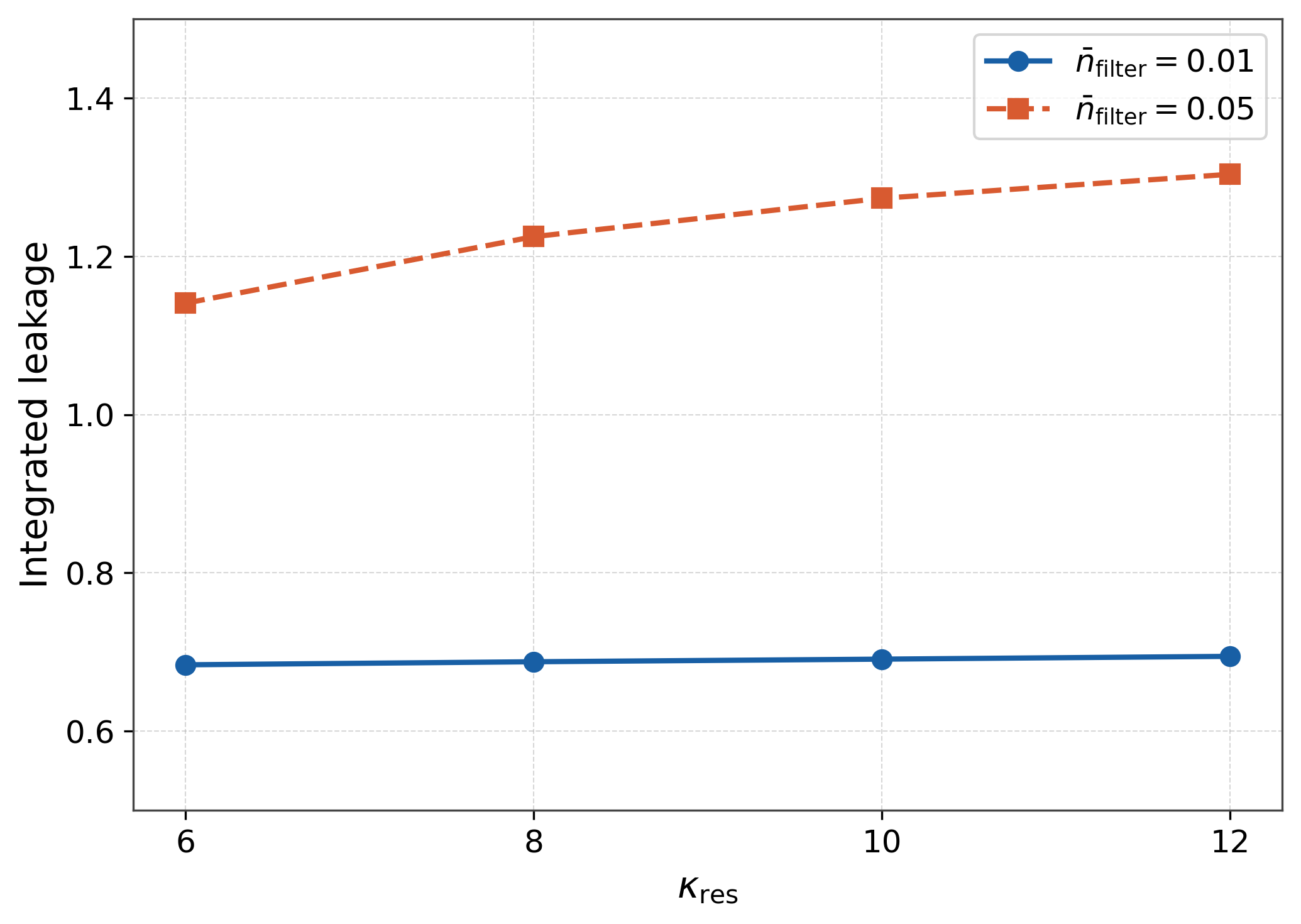}
    \caption{Integrated leakage as a function of resonator loss for two representative filter thermal occupations.}
    \label{fig:intkappa}
\end{figure}

Across the recorded sweep, moving from the low-loss settings labelled $3$--$4$ to the higher-loss settings labelled $8$--$10$ reduces the mean residual from $0.274\%$ to $0.164\%$ and the integrated leakage from $1.32$ to $1.12\,$ns. Within this effective model, the trend is consistent with faster removal of population through the dissipative channel. The corresponding residual and integrated-leakage trends are shown in Figs.~\ref{fig:reskappa} and~\ref{fig:intkappa}.

Changing the filter count from three to five modes reduces the mean residual from $0.189\%$ to $0.162\%$ and the integrated leakage from $1.21$ to $1.14\,$ns. In the seven-mode data, the residual does not continue to improve monotonically even though the integrated leakage decreases, indicating that the design problem is genuinely multi-objective rather than reducible to a single ``more modes is better'' rule.

\subsection{Ablation and control-parameter sensitivity}

The ablation provides the clearest example of why the endpoint residual and the transient leakage exposure are not interchangeable. With the auxiliary chain disabled, the control case eventually reaches $P_f(300\,\mathrm{ns})\simeq1.52\times10^{-8}$, but it spends much longer in $\lvert f\rangle$ and accumulates $\mathcal{L}_f=8.56\,$ns. Enabling the chain raises the long-time floor to $6.55\times10^{-4}$ while reducing the integrated leakage to about $0.78\,$ns. Using the sustained definition in Eq.~\eqref{eq:sustained}, the $10^{-2}$ threshold moves from $39.5$ to $9.76\,$ns and the $10^{-3}$ threshold from $59.3$ to $20.27\,$ns. So, depending on which quantity is used, the two configurations are ordered differently.

\begin{table}[t]
\centering
\caption{Ablation results at the representative operating point. Threshold times are sustained crossings.}
\label{tab:ablation}
\small
\begin{tabular}{lcccc}
\toprule
Configuration & Final $P_f$ & $\mathcal{L}_f$ (ns) & $t_s(10^{-2})$ & $t_s(10^{-3})$ \\
\midrule
Control & $1.52\times10^{-8}$ & 8.561 & 39.53 & 59.30 \\
Flux only & $1.52\times10^{-8}$ & 8.561 & 39.53 & 59.30 \\
Chain only & $6.55\times10^{-4}$ & 0.781 & 9.76 & 20.27 \\
Complete & $6.55\times10^{-4}$ & 0.783 & 9.76 & 20.27 \\
\bottomrule
\end{tabular}
\end{table}

\begin{figure}[t]
    \centering
    \includegraphics[width=\columnwidth]{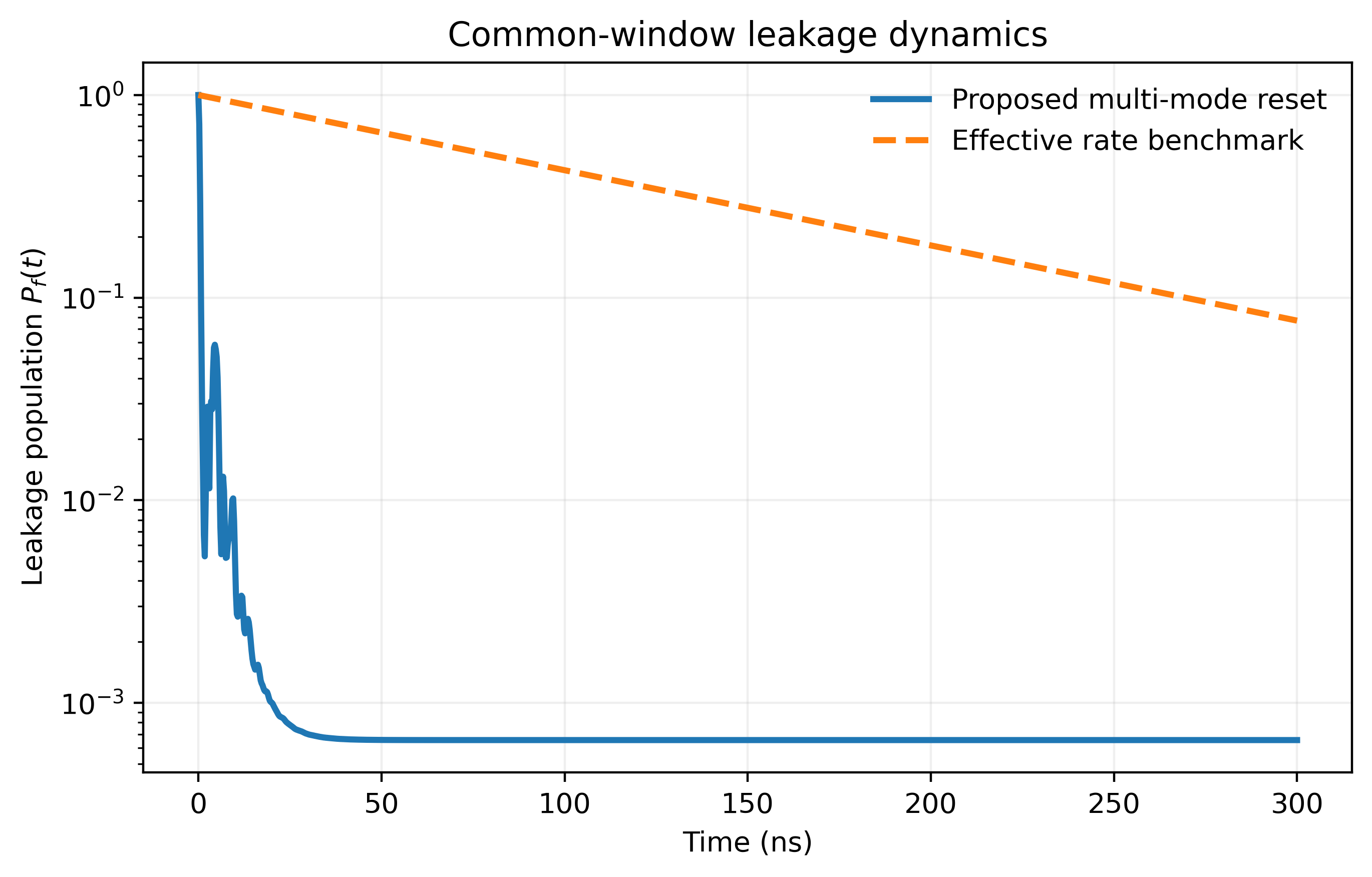}
    \caption{Common-window leakage dynamics for the complete configuration and the effective rate model parameterized by the $\lvert f\rangle$-state decay rate reported by Zhou \emph{et al.}~\cite{Zhou2021}. The benchmark is kinetic and is not a reproduction of the experimental device.}
    \label{fig:ratebenchmark}
\end{figure}

\subsection{Fabrication-disorder robustness}

Table~\ref{tab:disorder} summarizes the disorder sweep. The mean residual remains near $0.163\%$ across the tested range, and the integrated leakage remains close to $1.07\,$ns. The variation is small compared with the degradation produced by thermal occupation in the previous section.

\begin{table}[t]
\centering
\caption{Performance under random filter-frequency disorder.}
\label{tab:disorder}
\small
\begin{tabular}{ccccc}
\toprule
$\Delta$ (MHz) & Runs & Mean residual & Int. leak. (ns) & Final $\lvert g\rangle$ \\
\midrule
10 & 10 & $0.163\%$ & 1.071 & $96.05\%$ \\
20 & 10 & $0.163\%$ & 1.069 & $96.05\%$ \\
30 & 10 & $0.163\%$ & 1.066 & $96.05\%$ \\
40 & 10 & $0.162\%$ & 1.081 & $96.05\%$ \\
50 & 10 & $0.162\%$ & 1.087 & $96.06\%$ \\
60 & 10 & $0.164\%$ & 1.065 & $96.03\%$ \\
\midrule
Total & 60 & $0.163\%$ & $1.071\pm0.004$ & $96.05\%$ \\
\bottomrule
\end{tabular}
\end{table}

\begin{figure}[t]
    \centering
    \includegraphics[width=\columnwidth]{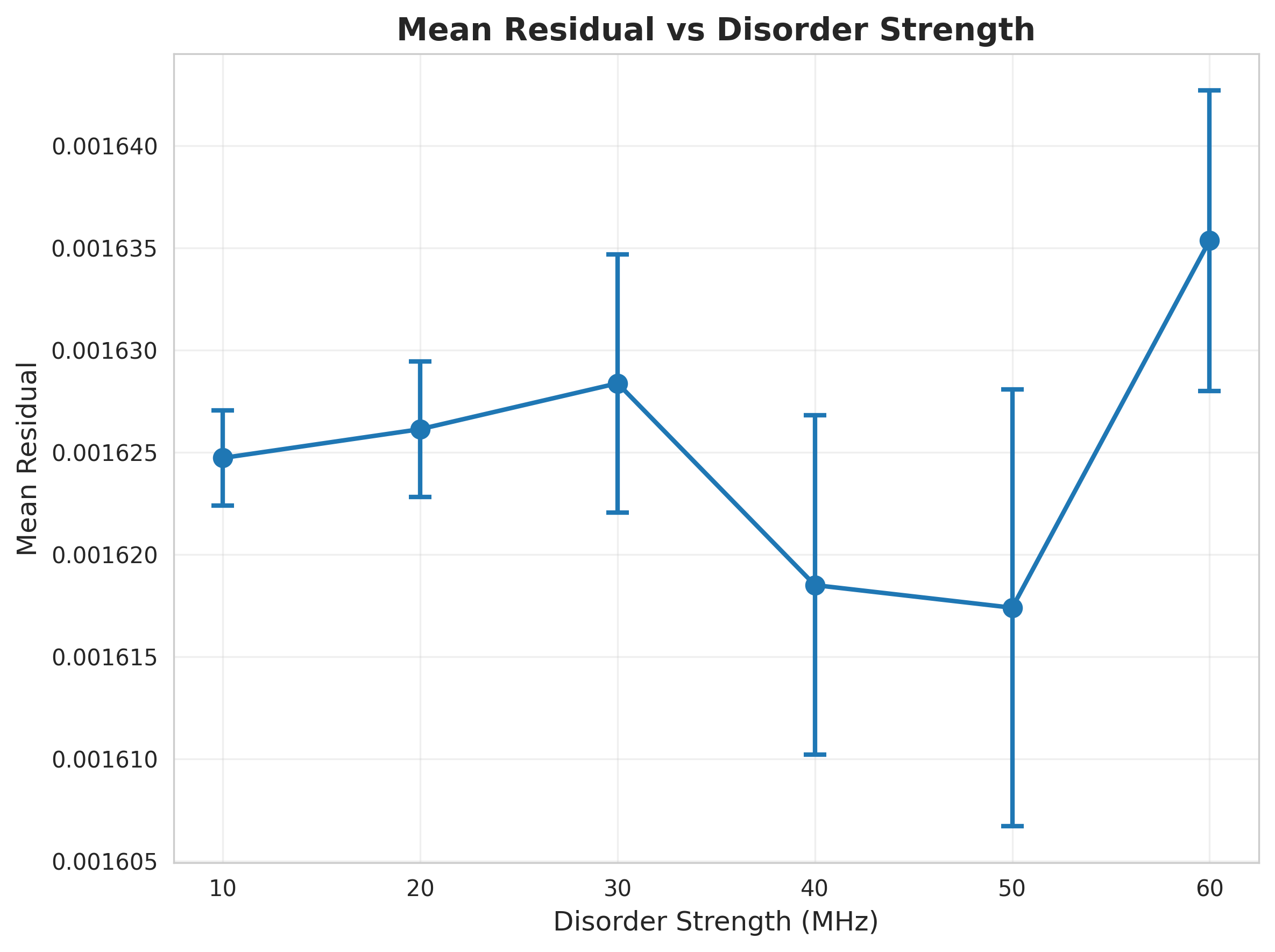}
    \caption{Mean residual population versus disorder strength. Error bars represent variation across independent frequency-disorder realizations.}
    \label{fig:disres}
\end{figure}

\begin{figure}[t]
    \centering
    \includegraphics[width=\columnwidth]{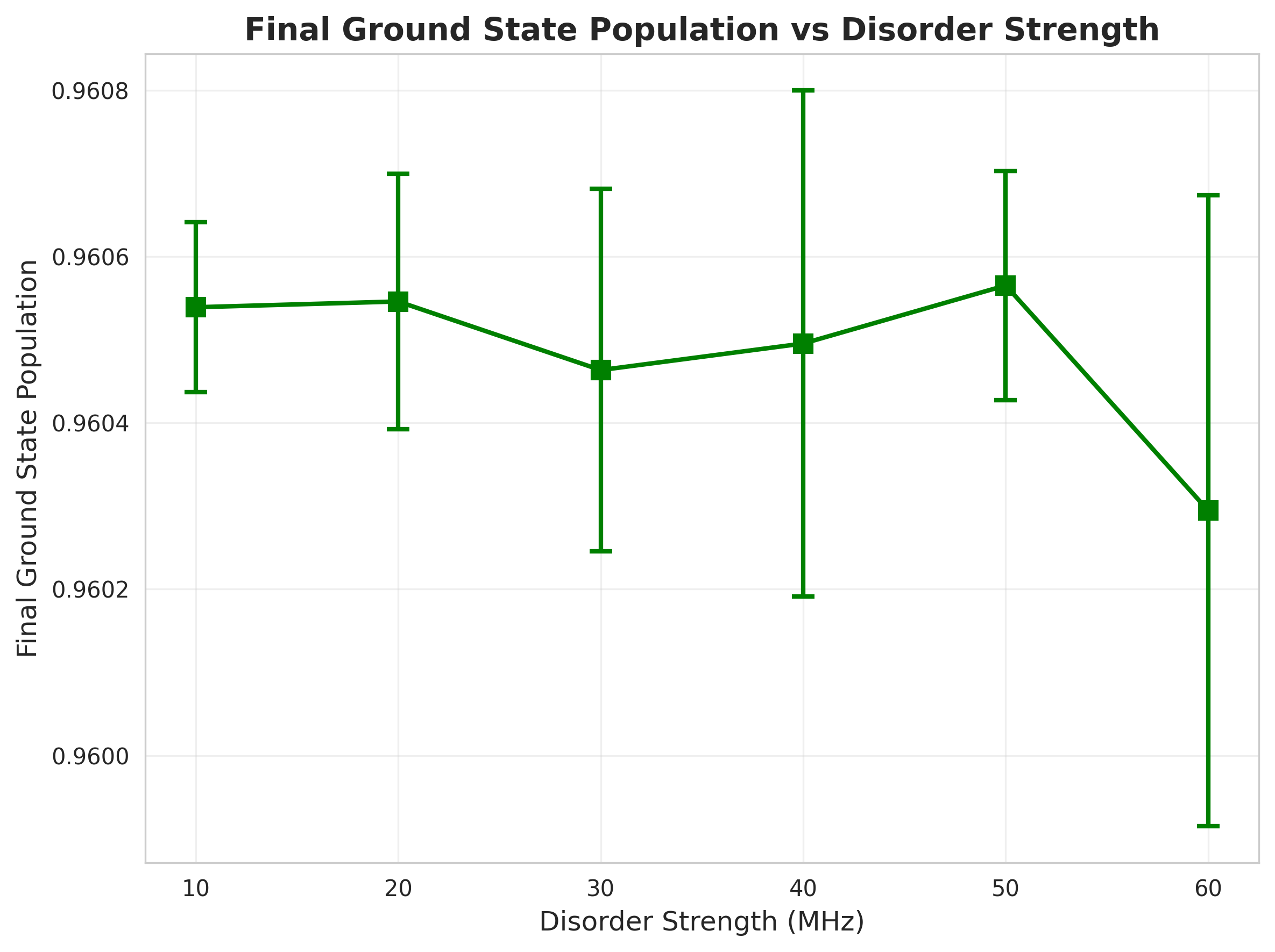}
    \caption{Final ground-state population versus disorder strength.}
    \label{fig:disground}
\end{figure}

\begin{figure}[t]
    \centering
    \includegraphics[width=\columnwidth]{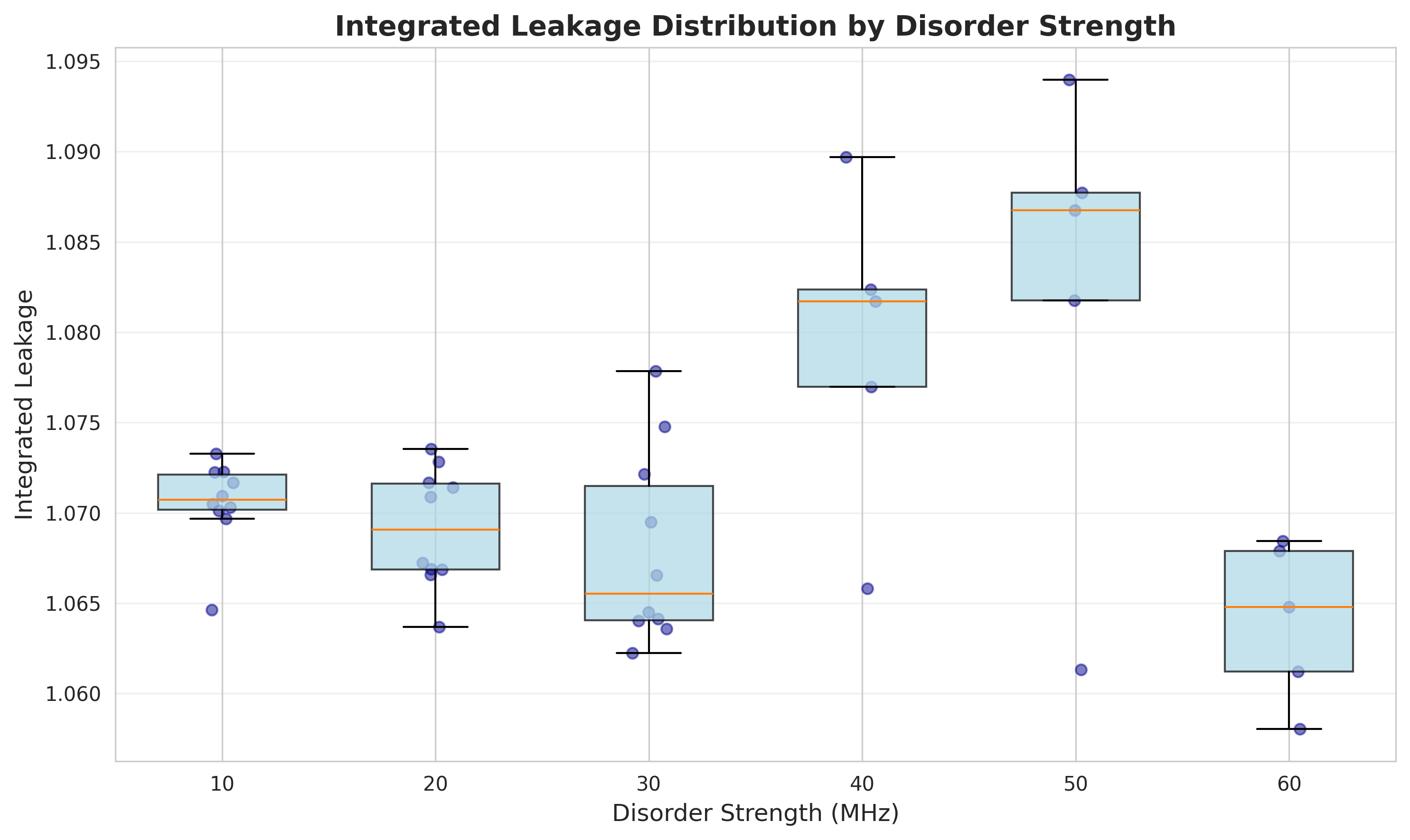}
    \caption{Distribution of integrated leakage for each disorder strength.}
    \label{fig:disint}
\end{figure}

\begin{figure}[t]
    \centering
    \includegraphics[width=\columnwidth]{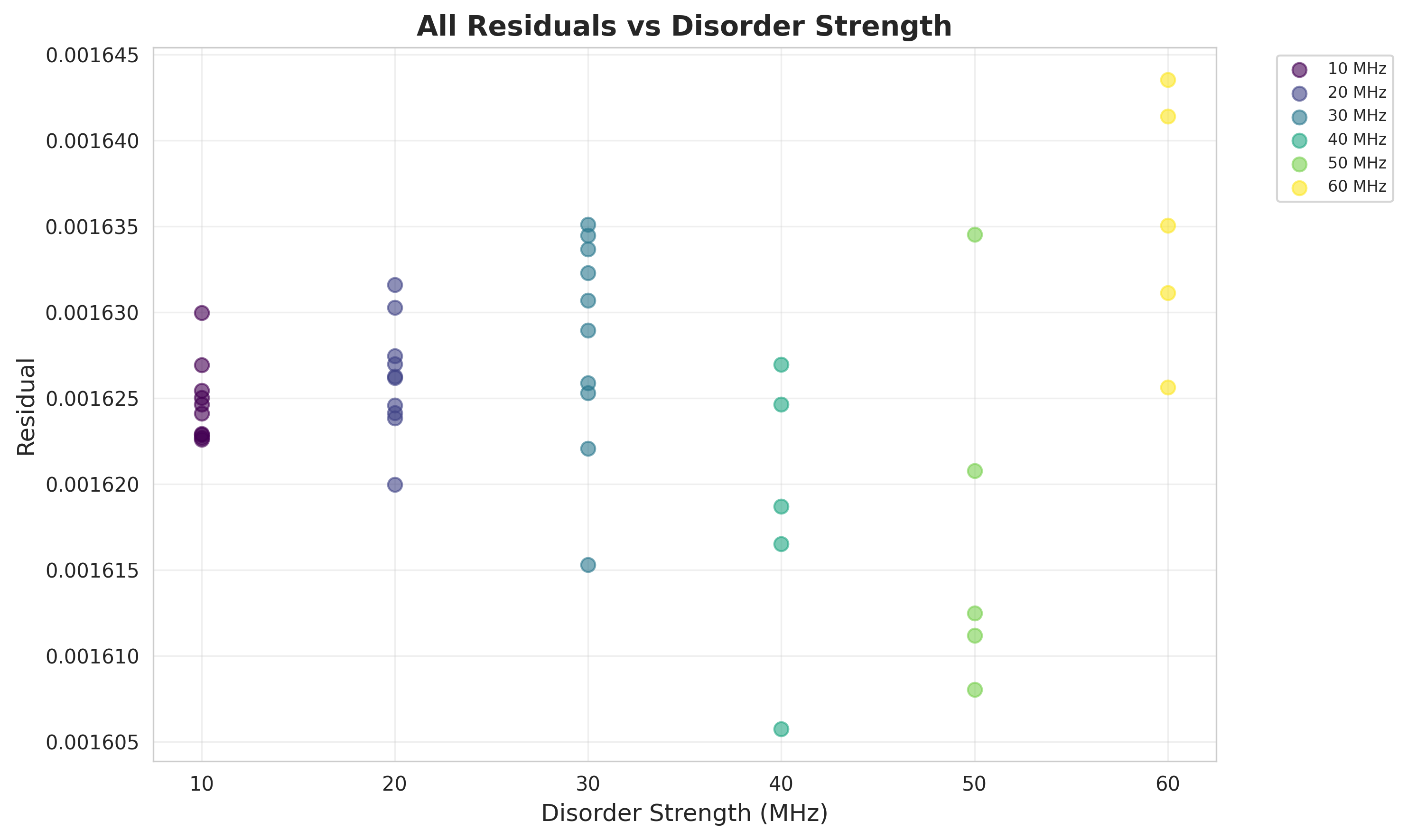}
    \caption{Residual populations from all fabrication-disorder realizations. No strong monotonic degradation is visible over the tested detuning range.}
    \label{fig:disall}
\end{figure}

\section{Discussion}

\subsection{What the simulations establish}

The main point of the simulations is not a single endpoint number. Instead, it is the observation that endpoint residual and transient leakage can lead to different conclusions for the same pair of reset configurations. In the ablation, the chain-free case reaches the smaller $P_f$ at $300\,$ns, while the chain-coupled case reduces accumulated leakage by about a factor of eleven and settles below the chosen thresholds much earlier. The sustained-threshold definition makes this distinction without being distorted by the early oscillations in $P_f(t)$.

The broader parameter sweeps help show the conditions under which this distinction becomes important. Finite thermal occupation is the dominant source of degradation in the sampled parameter space, whereas random auxiliary-mode detuning through $\pm60\,$MHz changes the reported averages only weakly. The flux-amplitude sweep produces little additional change at the representative operating point, so the rapid short-time leakage removal seen there is associated primarily with the auxiliary chain.

The literature-derived benchmark is intentionally simple. Its role is only to give an external timescale for the same metric. Over $300\,$ns, the fitted $\lvert f\rangle$ decay rate reported by Zhou \emph{et al.} implies $\mathcal{L}_f=107.99\,$ns in the effective cascade, compared with $0.783\,$ns for the complete configuration. The factor of $137.98$ belongs only to this effective-rate comparison; it is not an experimental performance ratio between two devices.

\subsection{Why integrated leakage can matter for QEC}

A simple way to interpret $\mathcal{L}_f$ is to imagine an error process that is active only while the transmon is in $\lvert f\rangle$. If its conditional rate is $\Gamma_L$ and is roughly constant during the reset window, the accumulated hazard is
\begin{equation}
\Lambda_L=\Gamma_L\int_0^{t_f}P_f(t)\,dt=\Gamma_L\mathcal{L}_f.
\label{eq:hazard}
\end{equation}
For a memoryless process,
\begin{equation}
P_{\mathrm{err}}^{(L)}=1-e^{-\Lambda_L}\simeq\Gamma_L\mathcal{L}_f,
\qquad \Gamma_L\mathcal{L}_f\ll1.
\label{eq:qecerr}
\end{equation}
This is an illustrative model rather than a universal leakage law; $\Gamma_L$ would depend on the circuit, code cycle, and microscopic coupling mechanism. Its purpose is simply to show why time spent in the leakage manifold can matter independently of the final population.

\subsection{Relation to prior reset experiments}

Direct comparisons between published reset protocols are difficult because the experiments do not use identical initial states, pulse definitions, resonator lifetimes, thermal environments, or stopping criteria. In particular, the $34\,$ns result reported by Zhou \emph{et al.} refers to the first minimum of an initially prepared $\lvert e\rangle$ population, whereas their two-tone $\lvert f\rangle$-depletion data evolve on a longer timescale and are summarized by the decay rates used here~\cite{Zhou2021}. Other demonstrations optimize different hardware and operational targets~\cite{Sunada2022,Yuan2023,Gu2025}. For that reason we do not present the literature as a single performance leaderboard.

The benchmark in Fig.~\ref{fig:ratebenchmark} is narrower: it asks what the published leakage-state decay rate implies for $P_f(t)$, $\mathcal{L}_f$, and threshold times when evaluated with the same initial state and analysis window. That makes the comparison useful for the metric while keeping the difference between the two physical systems explicit.

\subsection{Practical implementation considerations}

A hardware realization of this model would need a flux-control line for the modulation used here, a set of coupled auxiliary modes, and a controlled source of dissipation. Those ingredients are familiar in circuit-QED experiments, but the exact mapping between the effective model and a fabricated device still has to be worked out. The disorder study is encouraging only in the limited sense that, within the simulated range, moderate frequency offsets do not strongly change the reported metrics.

Thermalization remains the more important challenge. The strong dependence of residual population on $\bar n_{\mathrm{filter}}$ means that performance is likely to be set by the effective temperature and nonequilibrium photon occupation of the auxiliary modes. This points toward microwave filtering, infrared shielding, attenuation, and device-level thermal anchoring as experimentally relevant considerations.

\subsection{Limitations}

This is a numerical study, so the absolute values should be interpreted with that limitation in mind. The flux waveform is idealized; amplitude noise, phase noise, pulse distortion, and a real transfer function are not included. Crosstalk to neighboring qubits and parasitic modes are also absent. The auxiliary and resonator Hilbert spaces are deliberately small, and a systematic local-dimension convergence test has not yet been performed. In addition, the collapse operators are phenomenological effective channels rather than rates extracted from a particular fabricated circuit, and the disorder sweep changes mode frequencies without simultaneously varying every fabrication-sensitive coupling or linewidth.

The external benchmark has its own limitation. It uses fitted decay rates from Zhou \emph{et al.}, not their full driven qubit--resonator Hamiltonian, pulse sequence, thermal floor, resonator-depletion dynamics, or measurement model. The $137.98$ ratio therefore describes the effective-rate comparison defined in Eq.~\eqref{eq:ratebenchmark}; it should not be read as an experimentally demonstrated speedup. Finally, the ablation and flux-amplitude sweep were performed at one representative operating point, so the weak modulation dependence there should not be generalized to the full parameter space.

These limitations point to clear next steps: local-dimension convergence tests, experimentally calibrated dissipative rates, broader disorder sweeps that include coupling and linewidth variation, and multi-qubit simulations in which reset-induced leakage propagation can be measured directly.

\section{Conclusion}

Time-integrated leakage captures information that an endpoint population by itself does not contain. In the multi-mode open-system model studied here, enabling the auxiliary chain raises the asymptotic leakage floor but lowers $\mathcal{L}_f$ from $8.56$ to about $0.78\,$ns and shortens the sustained $10^{-2}$ threshold from $39.5$ to $9.76\,$ns. The same pair of configurations is therefore ranked differently depending on whether one cares about the final population or the transient leakage exposure.

The literature-derived rate model gives a second, external timescale for this observation. Over a common $300\,$ns window, the complete configuration gives $\mathcal{L}_f=0.783\,$ns, while an effective cascade parameterized by the $\lvert f\rangle$ and $\lvert e\rangle$ decay rates reported by Zhou \emph{et al.} gives $107.99\,$ns. The comparison is kinetic rather than device-level, but it shows how a published reset timescale can be translated into the same trajectory-based metric. Across the remaining simulations, thermal occupation is the dominant source of degradation, whereas the tested frequency disorder changes the results only modestly. For reset protocols intended to operate repeatedly inside QEC, it is therefore useful to report the endpoint residual together with a trajectory-sensitive quantity such as $\mathcal{L}_f$.

\section*{Acknowledgments}
We thank Prof. Ravishankar for helpful discussions. The open-system simulations were carried out with QuTiP using an ARTPARK-funded computing system.

\balance

\end{document}